\documentstyle[epsf,epsfig,aps]{revtex}
\draft
\input epsf
\begin{document}
\twocolumn[\hsize\textwidth\columnwidth\hsize\csname@twocolumnfalse%
\endcsname
\title{ONE OF POSSIBLE APPLICATIONS OF HIGH-Tc SUPERCONDUCTORS}
\author{A.V.NIKULOV}

\address{Institute of Microelectronics Technology and High Purity Materials, Russian Academy of Sciences, 142432 Chernogolovka, Moscow District, RUSSIA. E-mail: nikulov@ipmt-hpm.ac.ru}

\maketitle
\begin{abstract}
{Possible application of a result published recently is announced. According to this result the chaotic energy of thermal fluctuation can be transformed in electric energy of direct current in an inhomogeneous superconducting ring. Although the power of this transformation is very weak this effect has optimistic perspective of wide application because useful energy can be obtained from heat energy in the equilibrium state. It is proved that this process is possible in spite of the prohibition by the second law of thermodynamics. It is shown that the second law of thermodynamics can be broken in a quantum system. }
\end{abstract}

\

]
\narrowtext

\bigskip

\textbf{1. Introduction}

\bigskip

Superconductors are used first of all as the conductors with zero resistivity. But the superconductivity is not only the infinite conductivity. It is first of all a macroscopic quantum phenomena. The long-range phase coherence can exist in a superconducting state. The zero resistivity can be considered as one of the consequences of the phase coherence. A non-zero velocity of the superconducting electrons v$_{{\rm s}}$ can exist without a voltage in a state with long-range phase coherence because 
\begin{equation} \label{eq1} v_{s} = {\frac{{1}}{{m}}}(\hbar \nabla \varphi - {\textstyle{{2e} \over {c}}}A)
\end{equation}
 in this state. Here $\varphi $ is the phase of the wave function; A is the vector potential; m is the electron mass and e is the electron charge. Other consequences of the phase coherence are the Meissner effect [1], the quantization of the magnetic flux in a superconductor and the quantization of the velocity in a superconducting ring. In the relation
\begin{equation} \label{eq2} {\oint\limits_{l} {dlv_{s} = {\frac{{1}}{{m}}}(\hbar {\oint\limits_{l} {dl\nabla \varphi - {\frac{{2e}}{{c}}}{\oint\limits_{l} {dlA) = {\frac{{2\pi \hbar }}{{m}}}(n - {\frac{{\Phi} }{{\Phi _{0}} }})}} }} }} \end{equation}
obtained by the integration of (\ref{eq1}) along a closed path l, $n = {\textstyle{{1} \over {2\pi} }}{\oint\limits_{l} {dl\nabla \varphi} } $ must be an integer number since the wave function $\Psi $ = $\vert \Psi \vert $exp(i$\varphi $) of the superconducting electrons must be a simple function. $\Phi _{{\rm 0}} = \pi \hbar c/e =2.07 \ 10^{{\rm -} {\rm 7}} \ G \ cm^{{\rm 2}}$ is the flux quantum; $\Phi $ is the magnetic flux contained within the closed path of integration l. If the $\varphi $(r) function does not have a singularity inside l then n = 0. It is the cause of the Meissner effect. The $\varphi $(r) function can have a singularity only if a  nonsuperconducting hole is inside the superconductor. n can be a non-zero integer number in this case. If the superconductor thickness around the hole is big then a closed path l exists on which v$_{{\rm s}}$ = 0. According to (\ref{eq2}) $\Phi $ = n$\Phi _{{\rm 0}}$ in this case. In a opposite case of a superconducting ring with a narrow wall w the quantization of the velocity takes place. If the wall thickness is much smaller than the penetration depth of magnetic field w << $\lambda $ then LI$_{{\rm s}}$ << $\Phi _{{\rm 0}}$ and $\Phi  \approx BS$. The permitted values of the velocity in a homogeneous ring are 
\begin{equation} \label{eq3} v_{s} = \frac{\hbar }{Rm}(n - \frac{\Phi}{\Phi _{0}}) \approx  \frac{\hbar }{Rm}(n - \frac{BS}{\Phi _{0}}) \end{equation}
Here L is the inductance of the ring; I$_{{\rm s}}$ = sj$_{{\rm s}}$ =  s2en$_{{\rm s}}$v$_{{\rm s}}$ is the superconducting current in the ring; n$_{{\rm s}}$ is the density of superconducting pairs; l = 2$\pi $R; R is the ring radius; B is the induction of the magnetic field induced by an external magnet; S = $\pi $R$^{{\rm 2}}$ is the ring area. One of the consequences of the velocity quantization is the Little-Parks effect [2]. The $\vert $v$_{{\rm s}}\vert $ tends towards a minimum possible value. Therefore according to (\ref{eq3}) the energy of superconducting state of a ring (or a tube) and, as a consequence, the critical temperature of the ring 
\begin{equation} \label{eq4} T_{c} (\Phi ) = T_{c} [1 - \frac{\xi ^{2}(0)}{R^{2}}(n - \frac{\Phi}{\Phi_{0}})^{2}] \end{equation}
depend in a periodic manner on the magnetic field value, with period $\Phi _{{\rm 0}}$/S. The Little-Parks result [2] is the experimental evidence of the velocity quantization phenomena in superconductor. This phenomena is very important. The existence of forbidden states in superconducting ring means a possibility of a voltage induce by a change of the n$_{{\rm s}}$ value. Because the state with v$_{{\rm s}}$ = 0 is forbidden if the magnetic flux contained within the ring is not divisible by the flux quantum the superconducting current I$_{{\rm s}}$ = sj$_{{\rm s}}$ = s2en$_{{\rm s}}$v$_{{\rm s}}$ is not equal zero and changes with a n$_{{\rm s}}$ change. Consequently, according to the Faraday's law lE = - d$\Phi $/dt, a change of the superconducting pair density induces a voltage 
\begin{equation} \label{eq5} E = - \frac{1}{l}\frac{d\Phi}{dt} = - \frac{1}{l}[Ls2e\frac{\hbar }{Rm}(n - \frac{\Phi}{\Phi _{0}}){\frac{{dn_{s}} }{{dt}}} + L{\frac{{dI_{n}} }{{dt}}}]  \end{equation}
Here $\Phi $ = BS + LI$_{{\rm s}}$ + LI$_{{\rm n}}$ is the full magnetic flux inside the ring; I$_{{\rm n}}$ is the normal current in the ring;  I$_{{\rm n}}$ = lE/R$_{{\rm n}}$; R$_{{\rm n}}$ = l$\rho _{{\rm n}}$/s is the resistance of the ring in the normal state. After a sharp change of the n$_{{\rm s}}$ value the voltage decreases during the decay time L/R$_{{\rm n}}$. The n$_{{\rm s}}$ value can change in a consequence of a change of the ring temperature. Thus, a small superconducting ring can be used as a thermo-electric generator. In the present report the most interesting case, an induction of direct voltage in an inhomogeneous ring, is considered. 
\bigskip

\textbf{2. Inhomogeneous Superconducting Ring as Direct-Current Generator}
\bigskip
In a inhomogeneous superconducting ring with different density of superconducting pairs along the ring circumference the velocity has also different values. In the stationary state the normal current is equal zero and the superconducting current I$_{{\rm s}}$ = sj$_{{\rm s}}$ = s2en$_{{\rm s}}$v$_{{\rm s}}$ should be constant along the circumference. Consequently the velocity is higher in a section with lower n$_{{\rm s}}$ value. If a ring consists of two sections l$_{{\rm a}}$ and l$_{{\rm b}}$ (l$_{{\rm a}}$ + l$_{{\rm b}}$ = l = 2$\pi $R) with the density of superconducting pairs n$_{{\rm s}{\rm a}}$ and n$_{{\rm s}{\rm b}}$ the velocity (v$_{{\rm s}{\rm a}}$ and v$_{{\rm s}{\rm b}}$) and the superconducting current in the permitted state according to (\ref{eq2}) are equal
\begin{equation} \label{eq6} v_{sa} = \frac{2\pi \hbar } {m}{\frac{{n_{sb}} }{{(l_{a} n_{sb} + l_{b} n_{sa} )}}}(n - \frac{\Phi}{\Phi _{0}}) \end{equation}
\begin{equation} \label{eq7} v_{sb} =  \frac{2\pi \hbar } {m}{\frac{{n_{sa}} }{{(l_{a} n_{sb} +  l_{b} n_{sa} )}}}(n - \frac{\Phi}{\Phi _{0}}) \end{equation}
\begin{equation} \label{eq8} I_{s} = \frac{s4e\pi \hbar }{m}{\frac{{n_{sa} n_{sb}} }{{(l_{a} n_{sb} + l_{b} n_{sa} )}}}(n - \frac{\Phi}{\Phi _{0}})\end{equation}
Because the I$_{{\rm s}}$ value depends on the superconducting pair density of all ring section (see (\ref{eq8})) it can be changed by a n$_{{\rm s}}$ change in any ring section. The potential difference dV/dl appears at an inhomogeneous change of the n$_{{\rm s}}$ value along the ring circumference. At an alternate change of the n$_{{\rm s}}$ value the average of the potential difference over a long time $<dV/dl>$ is equal zero if n$_{{\rm s}}$ keeps a non-zero value in whole ring. But the $<dV/dl>$ value can be not equal zero if a ring section is switched iteratively from the superconducting to normal state and backwards.

According to (\ref{eq8}) I$_{{\rm s}} \ne $ 0 if $n_{{\rm s}{\rm a}} > 0$, $n_{{\rm s}{\rm b}} > 0$ and $\Phi $/$\Phi _{{\rm 0}}  \ne $ n. When (for example) the $n_{{\rm s}{\rm b}}$ value changes from $n_{{\rm s}{\rm b}} = n_{{\rm b}} > 0$ to n$_{{\rm s}{\rm b}}$ = 0 (at $n_{{\rm s}{\rm a}} = const > 0$) the potential difference V$_{{\rm b}}$ = R$_{{\rm b}{\rm n}}$I appears on the ring section l$_{{\rm b}}$ and the current I decreases exponentially from the initial value I = I$_{{\rm s}}$ during the decay time L/R$_{{\rm b}{\rm n}}$. Because the stationary state with I$_{{\rm s}}$ = 0 is forbidden at $n_{{\rm s}{\rm a}} > 0$, $n_{{\rm s}{\rm b}} > 0$ and $\Phi $/$\Phi _{{\rm 0}} \ne $ n the current in the ring increases to I $ \approx $ I$_{{\rm s}}$ = (s2eh/m)(n$_{{\rm s}{\rm a}}$n$_{{\rm b}}$/(l$_{{\rm a}}$n$_{{\rm b}}$+l$_{{\rm b}}$n$_{{\rm s}{\rm a}}$))(n - $\Phi $/$\Phi _{{\rm 0}}$) after the return of the l$_{{\rm b}}$ section to the superconducting state. A state with any integer number n is permitted. But the state with a minimum $\vert $ n - $\Phi $/$\Phi _{{\rm 0}}\vert $ value has the greatest probability. Therefore the average of I$_{{\rm s}}$ is not equal zero at $\Phi $/$\Phi _{{\rm 0}}  \ne $ n and $\Phi $/$\Phi _{{\rm 0}}  \ne $ n + 0.5. The average $<V_{{\rm b}}>$ induced by the switching of the ring section from the superconducting to normal state and backwards is not equal zero also in this case. 

If the energy difference between the lowest state with n and the next state with n+1 is bigger than k$_{{\rm B}}$T and $t_{{\rm n}} \gg  L/R_{{\rm b}{\rm n}}$, $t_{{\rm s}}\gg  L/R_{{\rm n}}$ then 
\begin{equation} \label{eq9}  < V_{b} > _{t} \approx {\frac{{s < n_{sb} >} }{{l_{b} n_{sa} + l_{a} < n_{sb} >} }}{\frac{{\Phi _{0}} }{{\lambda _{La}^{2}} }}(n - {\frac{{\Phi }}{{\Phi _{0}} }})Lf \end{equation}
Here t$_{{\rm s}}$ and t$_{{\rm n}}$ are duration of the superconducting and normal states of the l$_{{\rm b}}$ section; $\lambda _{{\rm L}}$ = (mc/4e$^{{\rm 2}}$n$_{{\rm s}{\rm a}}$)$^{{\rm 0}{\rm .}{\rm 5}}$ is the London penetration depth of the l$_{{\rm a}}$ section; f = 1/(t$_{{\rm s}}$ + t$_{{\rm n}}$) is the frequency of the switching from the superconducting to the normal state. This result published in [3,4] means that a small superconducting ring can be used as direct-current generator [5] in which the heat energy can be transformed to the electric energy of the direct current. 

If the temperature of the l$_{{\rm b}}$ section changes iteratively from T$_{{\rm m}{\rm i}{\rm n}}$ = T$_{{\rm c}{\rm b}}$ - $\Delta $T to $T_{{\rm m}{\rm a}{\rm x}} > T_{{\rm c}{\rm b}}$ a part of the heat energy Q spent during a time t can be transformed in the electric energy $E_{{\rm d}{\rm .}{\rm c}{\rm .}} = tW_{{\rm d}{\rm .}{\rm c}{\rm .}} = t<V_{{\rm b}}(t)I(t)>_{{\rm t}}$. The efficiency Ef = E$_{{\rm d}{\rm .}{\rm c}{\rm .}}$ /Q of such generator is 
\begin{equation} \label{eq10} Ef = {\frac{{R_{b}} }{{R_{b} + R_{l}} }}{\frac{{c_{s}} }{{c_{s} + c_{n} }}}{\frac{{\mu} }{{1 + \mu} }}(1 - {\frac{{T_{\min} } }{{T_{\max} } }}) \end{equation}
Here R$_{{\rm l}}$ is the resistance of a load; c$_{{\rm s}}$ = -T(dF$_{{\rm G}{\rm L}}$/dT) is the heat capacity of superconducting state; c$_{{\rm s}}$ + c$_{{\rm n}}$ = c$_{{\rm b}}$ is the whole heat capacity; $\mu $/(1+$\mu $) is the portion of the magnetic energy E$_{{\rm L}}$ = LI$_{{\rm s}}^{{\rm 2}}$/2 in whole change on the superconductor energy. It has the maximum value $\mu _{{\rm m}{\rm a}{\rm x}}$ = (32$\pi ^{{\rm 3}}$/$\kappa ^{{\rm 2}}$)(Ls/l$_{{\rm b}}^{{\rm 3}}$)(n-$\Phi $/$\Phi _{{\rm 0}}$)$^{{\rm 2}}$ at T = T$_{{\rm c}}$. $\kappa $ = $\lambda _{{\rm L}}$/$\xi $ is the superconductor parameter introduced in the Ginzburg-Landau theory. The maximum efficiency, at $R_{{\rm l}} \ll  R_{{\rm b}}$; $c_{{\rm n}} \ll  c_{{\rm s}}$ and $\mu  \gg  1$, is corresponded to the efficiency Ef = (1 - T$_{{\rm m}{\rm i}{\rm n}}$/T$_{{\rm m}{\rm a}{\rm x}}$) of the Carnot cycle considered first in 1824 year. 

Thus, the superconducting ring does not differ qualitatively from any conventional heat engine if the switching of the ring section from the superconducting to normal state is caused by the temperature change. Its power is very small. Therefore any application of such generator would be senseless if the switching could be in the consequence only of the T change. But this switching can take place at an unaltered temperature in a consequence of the thermal fluctuation. If the ring section l$_{{\rm a}}$ and l$_{{\rm b}}$ have different values of the critical temperature $T_{{\rm c}{\rm a}} > T_{{\rm c}{\rm b}}$ then at an unaltered T $ \approx $ T$_{{\rm c}{\rm b}}$, $n_{{\rm s}{\rm a}} > 0$ always whereas the l$_{{\rm b}}$ section is switched by fluctuation from the superconducting to normal state and backwards. The direct voltage (\ref{eq9}) should appear at both regular and chaotic switching. But the switching to the superconducting state should be simultaneous in whole section with lowest critical temperature. This simultaneous switching in the consequence of thermal fluctuation takes place in a mesoscopic section all sizes of which are not bigger than the superconducting coherence length $\xi $. At T $ \approx $ T$_{{\rm c}{\rm b}}$, the average resistance R$_{{\rm b}}$ of the l$_{{\rm b}}$ section $R_{{\rm b}{\rm n}} > R_{{\rm b}} = R_{{\rm b}{\rm n}}t_{{\rm n}}/(t_{{\rm s}}+t_{{\rm n}}) > 0$ because some times (t$_{{\rm n}}$) n$_{{\rm s}{\rm b}}$ = 0 (and therefore R$_{{\rm b}}$ = R$_{{\rm b}{\rm n}}$) and some times (t$_{{\rm s}}$) $n_{{\rm s}{\rm b}} > 0$ (and therefore R$_{{\rm b}}$ = R$_{{\rm b}{\rm n}}$). Consequently, a direct voltage can appear in a mesoscopic inhomogeneous superconducting ring at an unaltered temperature corresponded to the resistive transition of the section with lowest T$_{{\rm c}}$ [3,4].

\bigskip

\textbf{3. Possible Applications of Superconducting Ring System}

\bigskip

The direct voltage appeared in the consequence of the thermal fluctuation can be used. Therefore the superconducting ring can be considered as electric power source in which the chaotic energy of thermal fluctuation is transformed to the direct current energy. The power $W = <V_{{\rm b}}>^{{\rm 2}}R_{{\rm l}}/(R_{{\rm l}}+R_{{\rm b}})^{{\rm 2}}$ can develop across a load with the resistance R$_{{\rm l}}$. This power is obtained in a consequence of a regulating of the chaotic energy of the thermal fluctuation. 

\bigskip

3.1. LIMITING POWER OF FLUCTUATION ENERGY REGULATING

\bigskip

Fundamental relation for limiting power of this regulating is determined by the characteristic energy of fluctuation k$_{{\rm B}}$T and a time of a cycle which can not be shorter than $\hbar /k_{{\rm B}}T$ because of the uncertainty relation. Therefore $W < (k_{{\rm B}}T)^{{\rm 2}}/\hbar$  in any case. 

The power of one ring can not exceed 
\begin{equation} \label{eq11} W_{\max}  \approx {\frac{{\mu} }{{1 + \mu} }}{\frac{{16\pi k_{B}^{2}  T_{c}^{2} Gi}}{{h}}} \end{equation}
Here Gi = (k$_{{\rm B}}$T$_{{\rm c}}$/H$_{{\rm c}}^{{\rm 2}}$(0)l$_{{\rm b}}$s)$^{{\rm 0}{\rm .}{\rm 5}}$ is the Ginzburg number; H$_{{\rm c}}$(0) is the thermodynamic critical field at T = 0. This value is very small: at T$_{{\rm c}}$ = 100 K, (k$_{{\rm B}}$T$_{{\rm c}}$)$^{{\rm 2}}$2$\pi $/h $ \approx $ 10$^{{\rm -} {\rm 8}}$ Wt and 8$\pi $Gi$\mu $/(1+$\mu $) value can not exceed 1 in any ring. Therefore the power even of high-Tc superconducting ring can not exceed 10$^{{\rm -} {\rm 8}}$ Wt. In order to obtain an acceptable power a system with big number of the rings should be used. The power $W = N^{{\rm 2}}<V_{{\rm b}}>^{{\rm 2}}R_{{\rm l}}/(R_{{\rm l}}+NR_{{\rm b}})^{{\rm 2}}_{{\rm} }$ developed across the load R$_{{\rm l}}$ can be obtained by a system of N rings the sections of which are connected in series. This power has maximum value $W = N<V_{{\rm b}}>^{{\rm 2}}/4R_{{\rm b}}$ at R$_{{\rm l}}$ = NR$_{{\rm b}}$. Because the power $<V_{{\rm b}}>^{{\rm 2}}/R_{{\rm b}} <10^{{\rm -} {\rm 8}}$ Wt at T$_{{\rm c}}$ = 100 K we should used a system with 4 10$^{{\rm 8}}$ HTSC rings in order to obtain the power 1 Wt. 

Consequently in order to obtained the maximum possible power we should use HTSC rings with maximum possible Gi$\mu $/(1+$\mu $) value. 

\bigskip

3.2. REQUIREMENTS TO SUPERCONDUCTING RINGS AS DIRECT CURRENT GENERATOR

\bigskip

It is impossible now to give real optimum parameters of the ring system affording the maximum power of the fluctuation energy transformation. Additional theoretical and experimental investigations are necessary for this. I list only basic requirements which follow from the theoretical consideration. 

Because the fluctuation induce the voltage in a narrow temperature region near T$_{{\rm c}{\rm b}}$ and $\mu $ has maximum value at T = T$_{{\rm c}{\rm b}}$ all rings of the system should have the same critical temperature. A section of each ring should has a lower critical temperature T$_{{\rm c}{\rm b}}$ than T$_{{\rm c}{\rm a}}$ of other part of the ring. Sizes of this section should be enough small. They should not surpass considerably $\xi $/Gi$_{{\rm 3}{\rm D}}$. Here Gi$_{{\rm 3}{\rm D}}$ is the Ginzburg number of three-dimensional superconductor. The $\xi $/Gi$_{{\rm 3}{\rm D}}$ value of known HTSC is very small. Therefore methods of nano-technology are necessary for the making of the ring system. The l$_{{\rm a}}$ section with higher T$_{{\rm c}}$ can be longer considerably than l$_{{\rm b}}$. The $\mu _{{\rm m}{\rm a}{\rm x}}$ value increases with l$_{{\rm a}}$ because the ring inductance L is proportional to l. But the l$_{{\rm a}}$ should not be too long. At big l$_{{\rm a}}$/l$_{{\rm b}}$ value the transition to the superconducting state of the l$_{{\rm b}}$ section becomes first order if $\vert $ n - $\Phi $/$\Phi _{{\rm 0}}\vert $ is not small [3]. The fluctuation can not switch the l$_{{\rm b}}$ section from the superconducting to normal state and backwards if l$_{{\rm a}}$s is too big. I suppose the optimum diameter of HTSC ring should be approximately 1 $\mu $m. The system of 10$^{{\rm 8}}$ rings with diameter 1 $\mu $m can go in an area 1 cm$^{{\rm 2}}$. 

\bigskip

3.3.  POWER  SOURCE AND MICRO-REFRIGERATOR WITHOUT EXPENSE OF AN EXTERNAL ENERGY

\bigskip

Although the power of the proposed direct-current generator is very weak and its making is very difficult it has optimistic perspective of application. The proposed power source does not need an expense of an external energy. Moreover it can be used simultaneously as a micro-refrigerator [6]. The chaotic heat energy is transformed in the ordered electric energy in the proposed system of superconducting rings. The ordered energy turns back into the heat energy during an utilization. Therefore the proposed power source can work anyhow long time without an expence of any fuel. Such power source may be especially useful in self-cotained systems. 

\bigskip

\textbf{4. Why the chaotic heat energy can be ordered in a quantum system}

\bigskip

The perspective of considered application is connected with the possibility of the violation of the second law of thermodynamics which should take place [7] if the result published in [3,4] is correct. In a consequence of this circumstance most scientists can not believe that the result [3,4] can be correct. Below I try to prove that the proposed application of the HTSC rings can be real because the second law of thermodynamics can be broken in a quantum system. I explain also why the voltage with direct component can exist in the superconducting ring section.

\bigskip

4.1. A QUANTUM FORCE

\bigskip

According to (\ref{eq9}) the direct voltage $<E_{{\rm b}}> = <V_{{\rm b}}>/l_{{\rm b}}$ can be induced in the ring section by its iterative switching from the superconducting to normal state and backwards. Because 
\begin{equation} \label{eq12} E = - \nabla V - {\frac{{dA}}{{dt}}} \end{equation} 
and consequently   ${\oint\limits_{l} { < E > dl = < V_{a} > + < V_{b} > = - < {\textstyle{{d\Phi}  \over {dt}}} > = 0}} $  the direct voltage $<E_{{\rm a}}> = <V_{{\rm a}}>/l_{{\rm a}} = -<V_{{\rm b}}>/l_{{\rm a}}$ exists in the superconducting section l$_{{\rm a}}$ if it exists in the switched section l$_{{\rm b}}$. The superconducting pair in the l$_{{\rm b}}$ section should accelerate continually (because mdv$_{{\rm s}}$/dt = -eE) and consequently the result [3,4] can not be correct if only electric force F$_{{\rm e}}$ = -eE acts on electrons. But if we assumed this we would conclude that the Meissner effect is impossible. If 2mdv$_{{\rm s}}$/dt = -2eE then 2mdv$_{{\rm s}}$/dt + 2eE = d(2mv$_{{\rm s}}$ - 2eA/c)/dt = 0 and consequently the mv$_{{\rm s}}$- e$\Phi $/lc value should be constant in time. The magnetic flux should not change at the transition to the superconducting state. But the superconductivity is a macroscopic quantum phenomena. Therefore the superconducting state differs from the state with infinite conductivity. The superconducting pairs can accelerate against the electric force at the transition to the superconducting state as it takes place at the Meissner effect. Consequently an additional force can act on the electrons at the transition to the superconducting state. Because this force is caused by the quantization of the electron velocity I will call it as quantum force F$_{{\rm q}}$. 

The quantum force acts at the closing of the superconducting state in the ring. At the transition of the l$_{{\rm b}}$ section from normal n$_{{\rm s}{\rm b}}$ = 0 to superconducting state with n$_{{\rm s}{\rm b}}$ = n$_{{\rm s}{\rm a}}$ the velocity changes from v$_{{\rm s}}$ = 0 to v$_{{\rm s}}$ = (e/mcl)(n$\Phi _{{\rm 0}}$ - $\Phi $) and the magnetic flux inside the ring $\Phi $ changes from $\Phi $ = BS to $\Phi $ = BS + LI$_{{\rm s}}$. The mv$_{{\rm s}}$- e$\Phi $/cl value changes by (e/cl)(n$\Phi _{{\rm 0}}$ - $\Phi $) + (e/c)LIs = (e/cl)(1 + sL/$\lambda _{{\rm L}}^{{\rm 2}}$l)(n$\Phi _{{\rm 0}}$ - $\Phi $). Consequently at this transition $\int {dtF_{q} = \int {dt(m{\textstyle{{dv_{s}}  \over {dt}}} - {\textstyle{{e} \over {cl}}}{\textstyle{{d\Phi}  \over {dt}}})}}  $= (e/cl)(1 + sL/$\lambda _{{\rm L}}^{{\rm 2}}$l)(n$\Phi _{{\rm 0}}$ - $\Phi $). 

The quantum force F$_{{\rm q}}$ acts only at the transition from the normal to the superconducting state. At the transition of the l$_{{\rm b}}$ section to the normal state the v$_{{\rm s}}$ value in the l$_{{\rm a}}$ section decreases in a consequence of the electric force F$_{{\rm e}}$ = -eE. The electric field E is induced by the deceleration of electrons in the l$_{{\rm b}}$ section in a consequence of the energy dissipation. Thus, the superconducting pairs in the l$_{{\rm a}}$ section are accelerate by the quantum force F$_{{\rm q}}$ and are retarded by electric force F$_{{\rm e}}$ at the switching of the l$_{{\rm b}}$ section. They are not accelerated continually because $<F_{{\rm q}} + F_{{\rm e}}> = 0$. The $<F_{{\rm e}}>$ is not equal zero at the switching of the l$_{{\rm b}}$ section because $<F_{{\rm q}}>  \ne  0$. 

\bigskip

4.2. VIOLATION OF THE SECOND LAW OF THERMODYNAMIC

\bigskip

According to the Carnot's principle (1824 year), which may be considered as a first formula of the second law of thermodynamic, in the equilibrium state the heat energy can not be transformed to a work. The transformation of the fluctuation energy to electric energy of the direct current at an unaltered temperature considered above contradicts to this principle. It contradicts also to the Clausius's formula of the second law of thermodynamic (1850 year) if a temperature of the load is higher than the ring temperature. It contradicts to the Thomson's formula (1850 year) if the load is an electricmotor. Nevertheless I state that the results [3,4] is correct because the second law of thermodynamic can be broken in a quantum system. I understand it is very difficult convince anyone that this claim can be correct. Most persons are fully confident that the second law of thermodynamic can not be broken. Although few authors fight against this law (see [8]).Below I try to explaine because my claim is correct. 

The violation of the second law of thermodynamic means that the heat energy can be transformed to the work (in a regular energy) in the equilibrium thermodynamic state. That is the work is posssible without an expence of any fuel. Because the regular energy is transformed to the chaotic heat energy at the work, the work can be put out anyhow long time. The work is $A = \int {FdX} $. Here F is an abstract force; X is an abstract coordinate. Because dX = vdt, $A = \int {Fvdt} $. Consequently the violation of the second law of thermodynamic means that the average value $ < Fv > = {\int_{0}^{T} {Fvdt / T}} $ by the infinite time T can be not equal zero. Any closed thermodynamic system has come to the equilibrium state with maximum entropy during any time. Therefore the average $<Fv>$ should be considered in the equilibrium state. In the last centure the second law of thermodynamic was interpreted as the impossipility of any dynamic process in the equilibrium state. In this interpretation $<Fv> = 0$ because $F = 0$. 

In the beginning of our century physicists have understood that the dynamic processes take place in the equilibrium state in a consequence of the fluctuation. Therefore the fluctuation was considered as the breakdown of the second law of thermodynamic in that time [9]. But the fluctuation dynamic process is chaotic. Therefore $<Fv> = <F><v>$ and the work is equal zero if $<F> = 0$ or $<v> = 0$. In any conventional heat engine the force F and the velocity v are correlated. This correlation is achieved by an unequilibrium state. The entropy S = Q/T increases at the work of a conventional heat engine because the heat energy Q is transferred from a hot body with T$_{{\rm 1}}$ to a cold body with T$_{{\rm 2}}$ during this process and $\Delta S = Q/T_{{\rm 2}} - Q/T_{{\rm 1}} > 0$ if $T_{{\rm 1}} > T_{{\rm 2}}$. Numerous attempts to devise a correlated process with $<Fv>  \ne  <F><v>$ in the equilibrium state were unsuccessful for the present [8]. 

But the second law of thermodynamic can be broken at $<Fv> = <F><v>$ if $<F>  \ne  0$ and $<v>  \ne  0$. It is obvious that the motion should be circular in this case in order the work can be put out anyhow long time. In any classical (no quantum) system $<v> = 0$ because all states are permitted. The probability of a state P is proportional to exp-(E/k$_{{\rm B}}$T). The energy E of a state is function of v$^{{\rm 2}}$ in a consequence of the space symmetry. Therefore the probability of the state with v is equal the one with -v. Consequently $<v> = 0$ if all states are permitted. In a quantum state no all states are permitted. Therefore $<v>$ can be not equal zero in some quantum system. The superconducting ring with $\Phi $/$\Phi _{{\rm 0}}  \ne $ n and $\Phi $/$\Phi _{{\rm 0}}  \ne $ n + 0.5 is one of such system. At $\Phi $/$\Phi _{{\rm 0}}  \ne $ n and $\Phi $/$\Phi _{{\rm 0}}   \ne $ n + 0.5 the permitted states with the opposite directed velocity have different values of the kinetic energy E$_{{\rm k}{\rm i}{\rm n}}$ = mv$_{{\rm s}}^{{\rm 2}}$/2. For example at $\Phi $/$\Phi _{{\rm 0}}$ =1/4 the lowest permitted velocities in a homogeneous superconducting ring are equal v$_{{\rm s}}$ = -h/ml4 at n=0 and v$_{{\rm s}}$ = 3h/ml4 at n=1. The kinetic energy of these states differ in 9 times. Therefore the thermodynamic average of the velocity is not equal zero. 

The average force $<F>$ can be not equal zero at the switching of the ring section from the superconducting to normal state and backwards because the quantum force F$_{{\rm q}}$ acts at the closing of the superconducting state if $\Phi $/$\Phi _{{\rm 0}} \ne $ n. This force acts in the same direction (more exactly, has higher probability in one direction) if $\Phi $/$\Phi _{{\rm 0}}  \ne $ n + 0.5. This direction coincides with the velocity direction. Therefore $<F_{{\rm q}}><v> \ > \ 0$. This means that a part of the fluctuation energy can be ordered at the closing of the superconducting state in the ring. In order to switch the ring section to superconducting state the fluctuations should spend a energy E$_{{\rm k}{\rm i}{\rm n}}$ = sln$_{{\rm s}}$mv$_{{\rm s}}^{{\rm 2}}$/2 for the acceleration of the superconducting electrons and a energy E$_{{\rm L}}$ = LI$_{{\rm s}}^{{\rm 2}}$/2 for the induce of the magnetic flux $\Delta \Phi $ = LI$_{{\rm s}}$ of the superconducting current. This energy is no chaotic because the velocity has higher probability in one direction. The regulated energy is dissipated in the l$_{{\rm b}}$ section at its transition to the normal state. But we can use a part of this energy for a work if we put load a R$_{{\rm l}}$ on the voltage appeared on the l$_{{\rm b}}$ section as it is proposed above. 

Thus, the new application of HTSC proposed in the present work is real although it contradicts to the second law of thermodynamic. 

\bigskip

\textbf{5. Acknowledgments}

\bigskip

I am grateful Jorge Berger for the preprint of his paper and for stimulant discussion. This work is supported by the International Association for the Promotion of Co-operation with Scientists from the New Independent States (Project INTAS-96-0452) and National Scientific Council on Superconductivity. I thank these organizations for financial support.

\bigskip

\textbf{6. References}
\bigskip

1. Shoenberg, D. (1952) \textit{Superconductivity}, Cambridge. 

2. Tinkham, M. (1975) \textit{Introduction to Superconductivity}, McGraw-Hill Book Company, New-York.

3. Nikulov, A.V. and Zhilyaev, I.N. (1998) The Little-Parks effect in an inhomogeneous superconducting ring, \textit{J.Low Temp.Phys.} \textbf{112}, 227-236; \noindent http://xxx.lanl.gov/abs/cond-mat/9811148.

4. Nikulov, A.V. (1999) Transformation of Thermal Energy in Electric Energy in an Inhomogeneous Superconducting Ring, in \textit{Symmetry and Pairing in} \textit{Superconductors,} Eds. M.Ausloos and S.Kruchinin, Kluwer Academic Publishers, Dordrecht, 373-382; http://xxx.lanl.gov/abs/cond-mat/9901103.

5. Nikulov, A.V. (1999) A superconducting mesoscopic ring as direct-current generator, \textit{Abstracts of NATO ASI "Quantum Mesoscopic Phenomena and Mesoscopic Devices in Microelectronics"}  Ankara, Turkey, p.105-106 

6. Nikulov, A.V. (1999) A system of mesoscopic superconducting rings as a \noindent microrefrigerator, \textit{Proceedings of the Symposium on Micro- and Nanocryogenics}, Jyvaskyla, Finland, p.68.

7. Nikulov, A.V. (1999) Violation of the second thermodynamic law in a superconducting ring, \textit{Abstracts of XXII International Conference on Low} \textit{Temperature Physics,} Helsinki, Finland, p.498

8. Berger J. (1994) The fight against the second law of thermodynamics, \textit{Physics Essays} \textbf{7}, 281-296.

9. Smoluchowski, M. (1914) Gultigkeitsgrenzen des zweiten Hauptsatzes der Warmetheorie, in \textit{Vortrage uber kinetische Theorie der Materie und der Elektrizitat (Mathematische Vorlesungen an der Universitat Gottingen, VI}). Leipzig und Berlin, B.G.Teubner, p.87-105

\end{document}